\documentclass{ws-mpla}
\usepackage{bm}

\newcommand{\beq}{\begin{equation}}
\newcommand{\eeq}{\end{equation}}
\newcommand{\bea}{\begin{eqnarray}}
\newcommand{\eea}{\end{eqnarray}}
\newcommand{\D}{{\cal D}}
\newcommand{\Dslash}{\rlap{/}\kern-2.0pt D}

\begin{document}

\markboth{Matthew Wingate}
{${B}$ PHYSICS ON THE LATTICE: PRESENT AND FUTURE}

%
\catchline{}{}{}{}{}
%

\title{$\bm{B}$ PHYSICS ON THE LATTICE: PRESENT AND FUTURE}

\author{\footnotesize MATTHEW WINGATE}

\address{Institute for Nuclear Theory, University of Washington, Box 351550\\
Seattle, Washington, United States\\
wingate@phys.washington.edu}

\maketitle


\begin{abstract}
Recent experimental measurements and lattice QCD calculations 
are now reaching the precision (and accuracy) needed to
over-constrain the CKM parameters $\bar\rho$ and $\bar\eta$.
In this brief review, I discuss the current status of 
lattice QCD calculations needed to connect the experimental
measurements of $B$ meson properties to quark flavor-changing 
parameters.  Special attention is given to $B\to\pi\ell\nu$, 
which is becoming a competitive way to determine $|V_{ub}|$, 
and to $B^0-\overline{B^0}$ mixings, which now include 
reliable extrapolation to the physical light quark mass.
The combination of the recent measurement of the $B_s$
mass difference and current lattice calculations dramatically
reduces the uncertainty in $|V_{td}|$.
I present an outlook for reducing dominant lattice QCD 
uncertainties entering CKM fits,
and I remark on lattice calculations for other decay channels.

\end{abstract}

\ccode{PACS Nos.: 12.38.Gc, 13.20.He, 14.40.Nd}

\section{Introduction}

The main motivation for studying the properties of $B$ mesons 
is to understand how quarks change their flavor.  Is the
Cabibbo-Kobayashi-Maskawa (CKM) mechanism enough to describe
flavor-changing processes, or can we see the effects of new
physics?  

Even with the extraordinary effort made by the $B$ factories,
we presently see no cracks in the CKM model.  This allows us to
eliminate or tightly constrain many new physics models, 
afflicting them with a so-called flavor problem.  Continued agreement 
of experiment
with the CKM model will force us to explain the minimal
flavor violation.  Better yet, fissures will appear and possibly
favor one new model over another.  As usual in particle physics,
this type of indirect search complements
and guides direct searches.

Since quarks are confined inside hadrons, precise and
accurate theoretical calculations are necessary to connect
the experimental measurements of meson decay and mixing to
the fundamental quark couplings.  Many quantities
important for studying quark flavor are calculated using
lattice QCD (LQCD).  This brief review focuses on
calculations of $B$ decays and mixings.  Sec.~\ref{sec:lqcd} is
an overview of the methodology.

Semileptonic decays like
$B\to \pi\ell\nu$ and $B\to D\ell\nu$ provide information about
how the $b$ quark decays to a $u$ and $c$ quark by $W$ emission.
Since these decays proceed dominantly through tree diagrams, 
they allow one to determine the CKM matrix elements $|V_{ub}|$ and
$|V_{cb}|$, given clean experimental measurement of decay rates and
reliable theoretical calculation of hadronic matrix elements. 
$B\to\pi\ell\nu$ decays were hard to use to
determine $|V_{ub}|$ because of small branching fractions and 
large theoretical uncertainties.  In the past year both experiment
and theory have improved greatly, the latter through unquenched
LQCD calculations.  Section~\ref{sec:sl} reviews this
important development, gives an update on $B\to D\ell\nu$,
and discusses the difficulties and possibilities for lattice
calculations relevant to other semileptonic decays which might
also determine $|V_{ub}|$.

Mixing between neutral mesons proceeds through loop
diagrams to which new physics particles can contribute.
It is possible that new physics could be discovered by
assuming that only Standard Model particles contribute
and showing that the resulting CKM parameters
do not agree with the CKM parameters obtained from tree-level
decays.  Both $K^0 - \overline{K^0}$ and $B^0 - \overline{B^0}$ 
oscillations
provide important ways to search for deviations from the CKM
mechanism.  As in the case of semileptonic decays, calculations of
hadronic matrix elements are necessary in order to learn about
quark level processes from experimental measurements of
mass and lifetime differences.  Section \ref{sec:mixing} describes
recent progress made using LQCD in concert with new experimental
results.

Rare decays, like $B\to K^*\gamma$, occur through loop diagrams
called penguins.  In principle lattice QCD can also provide
reliable calculations of the relevant hadronic matrix elements
for these processes.  
Some technical difficulties facing LQCD contributions to 
rare decays are explained in Section \ref{sec:penguin}.

This review closes with the lattice prediction and 
subsequent experimental measurement of the $B_c$ mass (Sec.~\ref{sec:Bc})
and a few concluding remarks (Sec.~\ref{sec:concl}).


\section{Overview of Methodology}
\label{sec:lqcd}

To many people lattice QCD is a black box into which one
puts the strong coupling constant and quark masses 
and out of which one obtains the properties of hadrons.
If only that were so.  Those people who live inside the box
see LQCD is a method for numerically integrating
Euclidean spacetime path integrals, valid even when there
are no small parameters about which to form a perturbation
theory.  Communication between those living inside and outside
the box is imperative if progress is to be made.
Here I discuss some details of LQCD that will 
facilitate later discussions about recent progress and
future improvements and obstacles.  Interested readers
can find more detailed treatments elsewhere.\cite{LQCDmethods}

Expectation values of observables
are given by integrals over all possible field configurations.
For example the expectation value of operator $O$ is given by
\beq
\langle O \rangle ~=~
\frac{1}{Z}\int \!\D\psi \,\D\bar\psi \,\D U \; O[\psi,\bar\psi, U]\;
e^{-S[\psi,\bar\psi,U]}
\label{eq:Oexpect}
\eeq
where the $\psi,\bar\psi$ represent the quark, antiquark fields
and $U$ represents the glue field.  The action $S$ is the integral
over space and imaginary time of the QCD Lagrangian.
Lattice QCD provides a method for numerically
evaluating path integrals from first principles.  
The integral is made finite by
discretizing spacetime.  The presence of the damped exponential
in (\ref{eq:Oexpect}) means that only a small set of field configurations
will contribute to the path integral.  This is analogous to studying
a classical gas: the important configurations of particle positions
and velocities are a small fraction of the multitude of 
possible configurations, most of which give exponentially small
contributions to statistical traces over microstates.

The way that fermion antisymmetry is represented in path integrals
causes great difficulty for numerical methods.  The quark and
antiquark fields are valued over the field of anticommuting
complex numbers (Grassmann numbers) instead of the usual complex
numbers.  This is a formal construction which generally does not
permit numerical evaluation.  Luckily, if the action is
quadratic in the fermion fields, $S_f = \bar\psi\,Q\, \psi$,
the integral over fermion fields
can be done exactly, yielding a determinant, $\det Q$.
For example, the vacuum expectation value of a quark bilinear
is given by
\beq
\langle \bar\psi \,\Gamma \psi \rangle ~=~ \left. \int \! \D U
\; \frac{\delta}{\delta\bar\zeta}\,\Gamma\,\frac{\delta}{\delta\zeta}\;
e^{-\bar\zeta\,Q^{-1}[U] \,\zeta} \;\det Q[U]\; e^{-S_g[U]} 
\right|_{\zeta,\bar\zeta\to 0} 
\label{eq:quarkbilinear}
\eeq
where $\Gamma$ is a product of Dirac $\gamma$ matrices and
the $\zeta,\bar\zeta$ act as fermion sources.
The quantities calculated for $B$ physics results, 2- and 3-point
correlation functions, are straightforward generalizations of
(\ref{eq:quarkbilinear}).  The penalty we pay for this formal
integration is that a nonlocal updating algorithm must be used
to do importance sampling.  A necessary step involves inverting
a matrix which becomes singular as up/down quark masses
are decreased from artifically heavy to their physical values.

Writing (\ref{eq:quarkbilinear}) in detail allows us to 
define some terminology.
Note the 2 places the quark matrix $Q$ appears
in (\ref{eq:quarkbilinear}). These
correspond to 2 distinct steps in a lattice QCD calculation.
The determinant of $Q$ is included in the probability weight 
during the generation of typical, important glue field 
configurations.  The resulting collection of configurations
is then stored and later used to calculate ensemble averages
of operators.  In the case of fermionic operators, the inverse
of $Q$ is computed during this ensemble averaging.  The
term ``sea quarks'' refers to contributions of $\det Q$ to the
importance sampling and ``valence quarks'' refers to the 
fully dressed propagators $Q^{-1}$.  Logistically nothing
prevents one from using different quark mass values for
the sea and valence quark steps.  This procedure is called
``partial quenching'' and is useful since the generation
of configurations is much  more costly than the calculation
of quark propagators.\cite{Sharpe:2000bc,Cohen:1999kk}  Full QCD 
is in some sense a special case of partially quenched QCD
where $m_\mathrm{sea}=m_\mathrm{valence}$.

The last important point is that the
discretization $Q$ of the Dirac operator, $\Dslash + m$,
is not unique.  The choice of discretization has consequences
for the systematic uncertainties in numerical results
and for the computational expense.
One discretization, the improved staggered
quark action, permits numerical calculation with sea (and valence)
quark masses down to $1/10$ the strange quark mass.  (The
cost of the calculations increases rapidly as $m$ decreases.)
Ensembles of configurations which include effects of $2+1$
flavors of sea quarks using an improved staggered action, 
have been generated and made public by the MILC 
collaboration.\cite{Bernard:2001av}  The
effects of using light sea quarks are dramatic: cleanly 
computable quantities, many of which the quenched approximation 
got wrong by over 10\%, agree nicely with experiment.\cite{Davies:2003ik}
Calculations with other quark discretizations are desirable (1) to
further check the ``fourth-root'' algorithmic trick one needs with
staggered quarks to obtain $2+1$ sea quark flavors;\cite{Durr:2005ax} 
(2) to avoid the complications staggering creates in calculations
of light baryon properties.\cite{Golterman:1984dn,Tiburzi:2005is,Bailey:2005ss}
Unquenched calculations with
improved Wilson quarks and domain wall quarks are making steady
progress.  However, their advantages are tempered by having
to extrapolate using data with heavier quark masses, perhaps
without sufficient theoretical control.  Checks similar to
those done for staggered fermions are needed for
non-staggered actions to demonstrate that the extrapolations
to physical sea quark masses are under control.

\section{Semileptonic Decays and CKM Matrix Elements}
\label{sec:sl}

Semileptonic decays, where quark flavor is changed
by emission of a single $W$ boson, provide direct
access to the first 2 rows of the CKM matrix.  It is useful
to focus on these tree-level decays separately from
processes which begin at 1-loop level because new physics
particles generally should affect the latter more than
the former.

In order to use experimental measurements of semileptonic
decays to determine the CKM matrix elements, one needs
accurate calculations of the weak current between hadronic 
initial and final states.  In the next section the lattice calculations
needed for $|V_{cb}|$ and $|V_{ub}|$ 
are discussed (as are $|V_{cs}|$ and $|V_{cd}|$, briefly).
The mixing calculations relevant for 
$|V_{td}|$ and $|V_{ts}|$ are discussed in Sec.~\ref{sec:mixing}.
This brief review will not cover calculations for $|V_{us}|$ 
using lattice QCD, although there has been significant recent 
progress, using the leptonic decay 
constants\cite{Marciano:2004uf,Aubin:2004fs}
or the semileptonic form 
factor.\cite{Becirevic:2004ya,Okamoto:2004df,Dawson:2005zv,Tsutsui:2005cj} 
Finally, $|V_{ud}|$ is most precisely determined from nuclear and
neutron beta decays without needing lattice QCD.\cite{Blucher:2005dc}

\subsection{$B\to D^{(*)}\ell\nu$} 

Presently, the most precise determination of $|V_{cb}|$ comes
from inclusive measurements.\cite{Eidelman:2004wy} 
Improvements in lattice QCD calculations will help to reduce
the uncertainties in determinations from exclusive decays,
especially $B\to D^*\ell\nu$.  Regardless of which method
is most precise, cross-checks are important to truly over-constrain 
the CKM model.

The matrix element ${|V_{cb}|}$ may be determined from exclusive
semileptonic $B$ decays to the $D$ or $D^*$ by fitting the
experimental measurements of the differential partial decay width
to
\beq
\frac{d\Gamma}{dw}(B\to D^{(*)}\ell\nu) ~=~ 
\frac{G_F^2 |V_{cb}|^2}{48\pi^3} 
{\cal K}_{B\to D^{(*)}}(w) {\cal F}^{~2}_{B\to D^{(*)}}(w)
\eeq
where $w = v_B \cdot v_{D^{(*)}}$ is the velocity transfer
and ${\cal K}_{B\to D^{(*)}}(w)$ is a known kinematic function of $w$.
The shape of ${\cal F}_{B\to D^{(*)}}(w)$ near $w=1$ 
is given by dispersive bounds and HQET,\cite{Caprini:1997mu} 
and experimenters quote measured values for
${\cal F}_{B\to D^{(*)}}(1)|V_{cb}|$.  The overall normalization
can then be computed using lattice QCD.

The LQCD calculation can be made very precise by
constructing a double ratio of matrix elements, wherein
many uncertainties cancel.\cite{Hashimoto:1999yp}
Preliminary unquenched results give
${\cal F}_{B\to D}(1) = 1.074(18)(16)$, where
the first error are statistical and the second is
a combination of 1\% systematic errors.\cite{Okamoto:2004xg}
Combining this with the Winter 2006 experimental 
average\cite{Group(HFAG):2005rb} gives 
$|V_{cb}| = (39.7 \pm 4.2|_\mathrm{exp} \pm 0.9|_\mathrm{latt})
\times 10^{-3}$.  

While for $B\to D\ell\nu$ the dominant uncertainty 
lies on the experimental side, the 
experimental error is 2\% for $B\to D^*\ell\nu$.
Furthermore, Luke's
theorem says the heavy quark effective theory errors should 
also be smaller for the latter decay.  One hiccup is that the
LQCD extrapolation of ${\cal F}_{B\to D^*}(1)$ from the 
input light quark mass to the physical light quark mass passes
through the $D^*\to D\pi$ threshold.  Improved staggered
fermion calculations are poised to solve this problem for
2 reasons.  First, the calculations are able to be performed
within the chiral regime, so that heavy meson chiral perturbation
theory can provide a reliable formula for extrapolation.
Second, the presence of artificially heavy staggered pions
actually smoothes out the cusp in ${\cal F}_{B\to D^*}(1)$ at 
threshold as seen in staggered chiral perturbation 
theory.\cite{Laiho:2005ue}

\subsection{$B\to \pi\ell\nu$}

Because $\pi$ and $B$ have the same parity, only the
vector part of the weak current mediates $B\to\pi\ell\nu$.
The matrix element can be parameterized by 2 form factors:
\beq
\langle \pi(p_\pi) | V^\mu | B(p_B)\rangle \;=\;
f_+(q^2)\left( p_B^\mu + p_\pi^\mu - \frac{m_B^2 - m_\pi^2}{q^2} q^\mu
\right) + f_0(q^2) \frac{m_B^2 - m_\pi^2}{q^2} q^\mu \,.
\label{def:fplusf0} 
\eeq
The form factor $f_+(q^2)$ describes the momentum dependence
of the differential partial decay rate in the $B$ rest frame by
\beq
\frac{d\Gamma}{d q^2} ~=~ \frac{G_F^2}{24\pi^3} |\vec{p}_\pi|^3
|V_{ub}|^2 |f_+(q^2)|^2 \,.
\label{eq:Bpidiffrate}
\eeq
Experiments have measured the $B^0\to \pi^-\ell^+\nu$ 
branching fraction with impressive precision.  $|V_{ub}|$
is determined by integrating $f_+$ over $q^2$.  There is
presently a lower limit on $q^2$ below which the lattice
calculations incur large discretization errors: the $\pi$
should not have momentum larger than the inverse lattice
spacing.  Therefore, theoretical errors are minimized by
integrating over $16\,\mathrm{GeV}^2 \le q^2 \le q_\mathrm{max}^2$
and using the branching fraction with the same momentum 
cut.\cite{Flynn:1995dc}

\begin{figure}[t]
\vspace*{2pt}
\centerline{
\epsfig{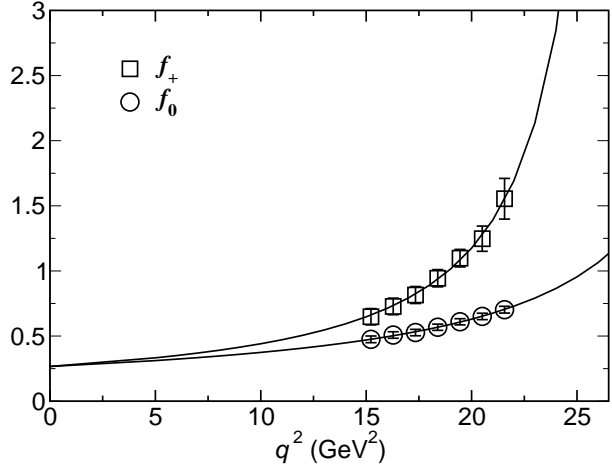}}
\vspace*{2pt}
\caption{\label{fig:f0pl}HPQCD results for the 
$B\to\pi$ form factors. The symbols
are lattice data after extrapolating the light quark mass 
to zero, and the curves are fits to the Ball-Zwicky parameterization.}
\end{figure}

Two groups are completing calculations of the $B\to\pi\ell\nu$
form factors.  They both use the MILC configurations, which
include effects of 2+1 quark flavors using the improved staggered
action.  The main difference is the heavy quark formulation:
either lattice NRQCD\cite{Gulez:2006dt} or the 
Fermilab method.\cite{Okamoto:2004df}  NRQCD is computationally
inexpensive but requires the heavy quark mass to be greater than
the inverse lattice spacing, precluding charm studies with NRQCD
on the MILC lattices.  The Fermilab method is valid for all 
quark masses.

One needs to fit the $q^2$ dependence of the form factors
to a functional form for two reasons.  First, to interpolate
the finite quark mass lattice data to fixed values of $E_\pi$, so
that the chiral extrapolation may be performed.\cite{Bowler:1999xn}
Second, to 
facilitate integration of $f_+$ over a range of $q^2$ corresponding
to the same range for a measured branching ratio, 
yielding a determination of $|V_{ub}|$.  
The commonly-used 3-parameter Be\'cirevi\'c-Kaidalov 
ansatz\cite{Becirevic:1999kt}
fits the HPQCD lattice data well only after extrapolation to the
chiral limit.  Instead we use the 4-parameter Be\'cirevi\'c-Kaidalov 
ansatz advocated by Ball and Zwicky.\cite{Ball:2004ye} 

The final HPQCD result is\cite{Gulez:2006dt}
\beq
\frac{1}{|V_{ub}|^2}\int_{16~\mathrm{GeV}^2}^{q^2_\mathrm{max}}
dq^2 \frac{d\Gamma}{dq^2} ~=~ 1.46 \pm 0.23\pm 0.27 ~\mathrm{ps}^{-1}
\label{eq:HPQCDintFF}
\eeq
where the first error is statistical plus chiral extrapolation
and the second is other systematics added in quadrature.  The dominant
source of systematic uncertainty is due to
truncating at 1-loop order the perturbative matching between the lattice 
NRQCD current and the continuum weak vector current.
 This agrees within errors
with the preliminary result using the Fermilab heavy quark action:
$|V_{ub}|^{-2}\int_\mathrm{16\,GeV^2}^{q^2_\mathrm{max}} dq^2 
(d\Gamma/dq^2) = 1.83 \pm 0.50$ ps${}^{-1}$.\cite{Okamoto:2004xg}
The dominant error in the latter calculation is due to 
discretization errors in their heavy quark action.
Taking current experimental results
for the $B^0\to \pi^-\ell^+\nu$ branching 
ratio\cite{Aubert:2005cd,Hokuue:2006nr,Group(HFAG):2005rb}
and $B^0$ lifetime\cite{Eidelman:2004wy}
along with the HPQCD result (\ref{eq:HPQCDintFF}) 
yields\cite{Gulez:2006dt}
\beq
|V_{ub}| ~=~ (4.22 \pm 0.30_\mathrm{exp} \pm 0.34_\mathrm{stat}
\pm 0.38_\mathrm{sys}) \times 10^{-3} \, .
\label{eq:VubHPQCD} 
\eeq

 With NRQCD for the heavy quarks, the leading uncertainty
comes from the perturbative matching of the lattice currents to
the continuum weak current.\cite{Gulez:2006dt}  
The matching is presently done at
1-loop level.  The 2-loop calculation is very involved and 
will require automated perturbation theory 
techniques.\cite{Trottier:2003bw}
With the Fermilab heavy quark formulation, there is a conserved
current which allows most of the matching to be determined 
nonperturbatively.  The remaining perturbative corrections are
tiny, so truncating at 1-loop level is sufficient until other
errors are decreased.  The leading uncertainty in this case
is the discretization error in the heavy quark formulation.  The
improved action has been worked out,\cite{Oktay:2003gk}
but an efficient implementation is still under development.

There are several efforts to reduce any systematic
uncertainties due to modeling the shape of the form factor.
The shape is strongly constrained by 
unitarity, analyticity
and heavy quark effective theory.\cite{Boyd:1994tt,Lellouch:1995yv}  
Plotted against an appropriate variable, the 
shape of $f_+$ is consistent with a linear fit.\cite{Mackenzie:fpcp2006}
Recent analyses\cite{Arnesen:2005ez,Becher:2005bg} make it apparent that 
LQCD calculations can reduce the uncertainty in $|V_{ub}|$ by more
precisely determining the form factor normalization.



We see that determinations of $|V_{ub}|$ from $B\to\pi\ell\nu$
decays are becoming competetive with the precision from 
inclusive determinations.\cite{Abe:2004zm}  Using exclusive
decays also has the advantage of avoiding the hypothesis
of quark-hadron duality which is necessary for inclusive
determinations and blunts tests of the CKM model.

\subsection{$D\to \pi \ell \nu$ and $D\to K\ell\nu$}

Even though $D$ decays are outside the scope of this brief
review, the recent calculations of semileptonic $D$ form factors
have implications for the $B\to\pi$ form factors, 
on top of providing determinations of $|V_{cd}|$ and 
$|V_{cs}|$.\cite{Aubin:2004ej}  The lattice calculations
are done with a subset of the same MILC configurations 
discussed above\cite{Bernard:2001av} and use the improved
staggered quark action in both the 
$B$ and $D$ correlation functions.\cite{Wingate:2002fh}  
The charm quark is treated using the Fermilab formulation, 
as in the Fermilab/MILC $B$ meson calculations.  
Given these similarities, the agreement between the lattice
calculation of $f_+(q^2)$ for $D\to K$ and the recent experimental
measurement\cite{Ablikim:2004ej,Link:2004dh}
testifies to the validity of the $B\to \pi$
form factor calculations.

\subsection{Other $b\to u\ell\nu$ Decays and Associated Challenges}

Exclusive determinations of $|V_{ub}|$ from other exclusive $B$
decays $B\to\pi\ell\nu$ are also valuable.  
Every independent measurement has the potential to
play a role in the quest to see
new physics through over-constraining the CKM parameters.
Even within the Standard Model, it is conceivable that another decay
channel will lead to a more precise determination than
$B\to\pi\ell\nu$, provided that lattice QCD can meet some
challenges.  

Experiments show the $B\to\rho\ell\nu$ branching fraction to be
twice as big as to $\pi\ell\nu$, so its precision is easier to
reduce. However, 
to get an accurate $|V_{ub}|$ the theoretical uncertainties in the
form factors must be under control; presently they
are not.\cite{Bowler:2004zb}  The main open problem is the
exptrapolation of lattice data obtained with quark masses
where the $\rho$ is stable through $\rho\to\pi\pi$ 
threshold.
The firm theoretical ground provided by chiral perturbation theory 
for $B$ and $\pi$ properties becomes quite muddy underneath
the $\rho$.

Experiments have measured $B\to\eta\ell\nu$.\cite{Athar:2003yg}  
The lattice calculation with the $\eta$ suffers because it is 
partially a flavor singlet.  There are disconnected 
contributions to the correlation functions
where the valence $\bar{q} q$ pair are annihilated then recreated.
These are notoriously hard to calculate due to poor
signal-to-noise.  Nevertheless, these contributions are
important to the $\eta$ propagator due to the anomaly.

The situation may not be as glum for 
$B\to\omega\ell\nu$,\cite{CM:JS:disc2005} which has
recently been observed.\cite{Schwanda:2004fa}  
The $\omega$ is a narrow resonance in 
contrast to the $\rho$.
Despite being a flavor singlet, violations of
the OZI rule are expected to be smaller for the vector $\omega$
than for the pseudoscalar $\eta$, since the anomaly affects only the 
latter.  LQCD calculations of flavor singlet meson masses
support this expectation so far.\cite{Isgur:2000ts,McNeile:2001cr}
We should explore in more depth the reliability of calculating the 
properties of the $\omega$ on the lattice and see how far we can
push a calculation of the $B\to\omega$ form factors.
If we can truly neglect disconnected contributions to
the $\omega$ propagator, then the lattice calculation 
of $B\to\omega$ form factors would be identical to $B\to\rho$,
with less headache regarding the meson's width.
Disconnected contributions are routinely ignored in LQCD
calculations of the $\phi$ meson mass, for example.

\section{Mixing and the Search for New Physics}
\label{sec:mixing}

Studying the oscillations of neutral mesons has the potential
to expose physics beyond the Standard Model in a way that
the semileptonic decays do not.  Any new physics entering the
semileptonic decays would be a small contribution to the
tree-level standard model decays. In contrast
oscillations begin at one-loop level in the standard electroweak 
theory, and new
particles can enter directly in those loops.  The leading
uncertanties so far are theoretical.  Nevertheless, with increased 
precision in lattice calculations and with the new
measurement of $B_s^0 - \overline{B_s^0}$ oscillations,
the potential to see a breakdown of the CKM model is realistic,
even if unrealized presently.
This section summarizes the most recent lattice QCD results
and prospects for further improvement.

Before turning to the $B^0$ mixings, 
let us consider $K^0$ and $D^0$ mixings.
Neutral kaon mixing is another important window into flavor physics
which relies on lattice QCD.  The theoretical error presently
dominates that constraint on $(\bar\rho,\bar\eta)$.  
This situation should improve
as we see more unquenched calculations of the hadronic matrix element
parametrized by $B_K$.\cite{Dawson:2005za}

$D^0 - \overline{D^0}$ mixing proceeds very slowly compared to mixing 
the $K$ and $B$ systems.
In the Standard Model, this is because the top quark mass, more
than 40 times $m_b$, enhances the $\Delta B = 2$
and $\Delta S=2$ box diagrams compared to $\Delta C=2$.
Of course, rare processes are just the kind one wants to study
to discover new physics.  As soon as precise experimental measurements
can be made, lattice QCD would be valuable in connecting the meson mass 
and width differences to models of quark flavor-changing interactions.
The relevant matrix elements were calculated some time 
ago\cite{Gupta:1996yt,Lellouch:2000tw} and should be updated using 
unquenched configurations and the Fermilab formulation for the charm quark.

\subsection{$B^0 - \overline{B^0}$}

In the Standard Model $B^0 - \overline{B^0}$ mixing proceeds
via 1-loop box diagrams, with the dominant contribution coming
from a top quark in the loop.  Measurements of the oscillation
frequency $\Delta m_d$ can  be used to constrain the CKM
matrix element $|V_{td}|$.
The relevant hadronic matrix element is 
conveniently parameterized as
\beq
\langle \overline{B^0} | (\overline{b}d)_{V-A}(\overline{b}d)_{V-A}
|B^0 \rangle \equiv \frac{8}{3} f_{B}^2 m_{B}^2 \mathrm{B}_{B} \, .
\label{eq:LLop}
\eeq
This is helpful for two reasons.  The leptonic decay constant $f_B$ 
is simpler to compute in lattice QCD. In addition chiral perturbation theory
shows that $f_B$ is more sensitive to the quark mass than 
$\mathrm{B}_B$, the ratio of the full matrix element to its value 
in the vacuum saturation approximation.

\begin{figure}[t]
\vspace*{2pt}
\centerline{
\epsfig{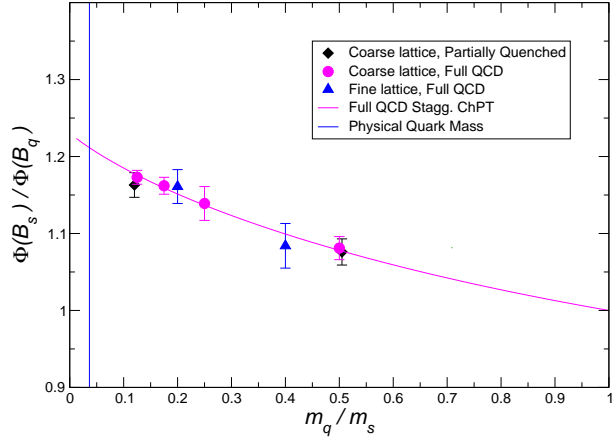}}
\vspace*{6pt}
\caption{\label{fig:phirat}Ratio of $f_{B_s}/f_{B_q}$ (times
$\sqrt{m_{B_s}/m_{B_q}}$) vs.\ mass of the light quark 
(in units of the physical strange quark mass).}
\end{figure}

As discussed above, the LQCD calculations are done with
quark masses $m_q$ larger than the physical up/down quark mass.
Let's denote the corresponding decay constant by $f_{B_q}$.
In the past, the extrapolation of $f_{B_q}$ to the physical
light quark mass was done phenomenologically.  Since the lattice
data for $f_{B_q}$ are linear in the regime $m_s/2 < m_q < 2 m_s$,
a linear fit was used for the extrapolation. Chiral perturbation
theory gives the $m_q$ dependence of $f_{B_q}$ including
logrithmic curvature; however the 
theory must be used in the small $m_q$ regime, approximately
$m_q \le m_s/2$.  The use of improved staggered quarks for
the fermion determinant and for the light quark propagator
allows LQCD calculations to be performed inside this chiral
regime.

The HPQCD collaboration has used the MILC lattices discussed above
to calculate $f_{B_q}$ vs. $m_q$ (see Fig.~\ref{fig:phirat}).
The data are fit using an extension of chiral perturbation theory
suitable for staggered fermion calculations.\cite{Aubin:2005aq}
Our result for $f_B$ is
$216 \pm 9_\mathrm{stat} \pm 20_\mathrm{sys} ~\mathrm{MeV}$.\cite{Gray:2005ad}
The leading uncertainty is due to the perturbative matching
between the currents in the heavy quark effective theory 
(lattice NRQCD) and the continuum
weak axial vector current, which has been done at 1-loop
order.\cite{Gulez:2003uf}  As in the case of the $B\to\pi$ form
factors, reducing this uncertainty requires a difficult
2-loop calculation.

Very recently the Belle experiment reported a result
$f_B = 176 ({}^{+28}_{-23})_\mathrm{stat}({}^{+20}_{-19})_\mathrm{sys}$ MeV
from a first measurement of $B^-\to \tau^- \bar\nu_\tau$ combined
with $|V_{ub}|$ from $B\to X_u\ell\nu$.\cite{Ikado:2006un,Group(HFAG):2005rb}
The confirmation of the LQCD prediction directly speaks to the 
validity of LQCD calculations for $B$ meson properties, in particular
the use of NRQCD heavy quarks, staggered light quarks,
and staggered chiral perturbation theory.

In order to further judge the reliability of the lattice QCD calculation
of $f_B$, it is useful to look at the $D$ decay constant.
A lattice calculation, which differed from the $f_B$ calculation only
in heavy quark action, recently predicted $f_D$ to be 
$201\pm 3_\mathrm{stat} \pm 17_\mathrm{sys}$ MeV,\cite{Aubin:2005ar}
much more precise than the 20\% experimental uncertainty at the 
time.\cite{Bonvicini:2004gv}
Subsequently, CLEO-c found
$f_D^\mathrm{expt} = 223\pm 17_\mathrm{stat} \pm 3_\mathrm{sys}$ 
MeV from $D^+\to\mu^+\nu$.\cite{Artuso:2005ym} 
This successful prediction by lattice QCD using the same
MILC configurations, light quark action, and staggered chiral
perturbation theory provides a significant test of the $f_B$ 
calculation.

The remainder of the hadronic matrix element,
$\mathrm{B}_B$, remains to be calculated on the
MILC lattices, although preliminary efforts have begun.\cite{Gray:2004hd}
The state-of-the-art is still the JLQCD calculation done with
improved Wilson fermions.\cite{Aoki:2003xb} The same systematic
uncertainties affect their result as the HPQCD result for $f_B$:
truncating the perturbative and heavy quark expansions.  
Since the chiral extrapolation for $\mathrm{B}_B$ is milder than 
for $f_B$, it is reasonable to expect their result to be close to what 
will be obtained with lighter quark mass computations.  Combining
the 2 groups' results one obtains 
$f_B\sqrt{\hat\mathrm{B}_B} = 244\pm 26$ MeV.\cite{Okamoto:2005zg}
This reduces the uncertainty down to 11\% from the previous
world average $f_B\sqrt{\hat\mathrm{B}_B} = 214\pm 38$ 
MeV.\cite{Hashimoto:2004hn}

\begin{figure}[t]
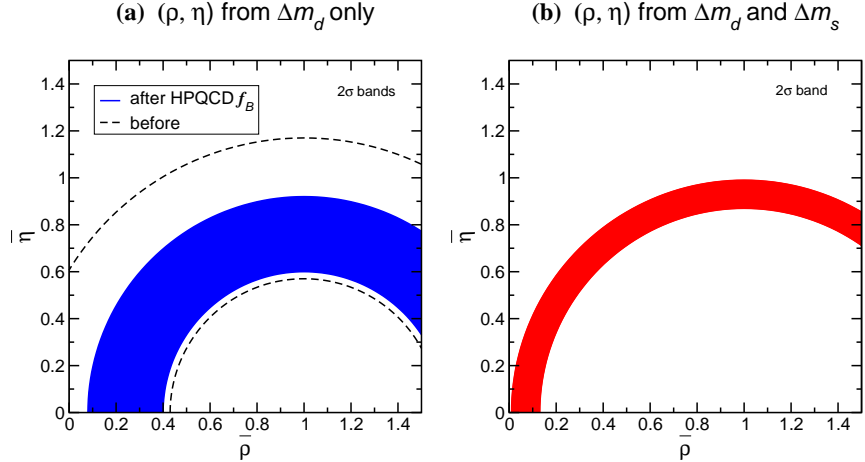

\vspace*{2pt}
\centerline{
\epsfig{file=VtdBd.eps,width=55mm}~~~\epsfig{file=VtdBdBs.eps,width=55mm}}
\caption{\label{fig:Vtdconstraint}Constraints (at the
$2\sigma$ level) on $|V_{td}| = A\lambda^3\sqrt{(1-\bar\rho)^2
+ \bar\eta^2}$.
(a) Old and new constraints from $\Delta m_d$.  
(b) Constraint from $\Delta m_d/\Delta m_s$ using the new CDF 
measurement and recent LQCD improvements.}
\end{figure}

Fig.~\ref{fig:Vtdconstraint}a shows the
constraint in the $\bar\rho-\bar\eta$ plane from $\Delta m_d$.  
The improvement indicated by the 2 bands is
achieved by reducing the long extrapolation of $f_B$
in light quark mass.

For physics beyond the standard Model, $B^0-\overline{B^0}$ oscillations
may be governed by 4-quark operators with different structure than in
(\ref{eq:LLop}); this is certainly the case for supersymmetric
extensions of the Standard Model.\cite{Gabbiani:1996hi}  
Therefore, matrix elements of all 5 possible operators should
be computed using unquenched lattice QCD, as has been done in
the quenched approximation.\cite{Becirevic:2001xt}

Given the difficulty of 2-loop calculations in lattice
perturbation theory, compounded by complexity of heavy quark
lattice actions, a fully nonperturbative method would be most
welcome.  Several such approaches are being 
investigated.\cite{Guagnelli:2002jd,Heitger:2003nj,Lin:2005ze,Palombi:2006pu}
Only time will tell which of these approaches will first
surpass in precision the methods used in present calculations 
and when that might occur.

\subsection{$B_s^0 - \overline{B_s^0}$}

Unlike leptonic $B$, $D$, and $D_s$ decays, which are
merely helicity-suppressed, leptonic $B_s$ decays are forbidden
in the Standard Model since they would proceed through
a flavor-changing neutral current.  Nevertheless, the 
decay constant $f_{B_s} \equiv \langle 0| A_0 | B_s\rangle/m_{B_s}$
is finite and approximately equal to $f_B$.  Again, it provides
a useful parameterization of the
hadronic matrix element relevant for $B_s^0 - \overline{B_s^0}$ mixing,
as in (\ref{eq:LLop}).

Very recently, the D0 experiment reported a 2-sided bound on the 
oscillation frequency
$17\;\mathrm{ps}^{-1} < \Delta m_s < 21\;\mathrm{ps}^{-1}$
at the 90\% confidence level,\cite{Abazov:2006dm} and
the CDF collaboration reported a measurement
of $\Delta m_s = 17.3 {\;}^{+0.4}_{-0.2}\;\mathrm{ps}^{-1}$.\cite{CDF:fpcp2006}
Some theoretical uncertainties cancel in ratio of
the neutral $B_s$ and $B$ oscillation frequencies, so a theoretical
calculation of $f_{B_s}^2 \mathrm{B}_{B_s}/f_B^2 \mathrm{B}_B$
can now provide a much tighter constraint on $|V_{td}|$ than using
only nonstrange $B$ oscillations.  

The HPQCD result for $f_{B_s}$ is
$260 \pm 7_\mathrm{stat} \pm 28_\mathrm{sys} ~\mathrm{MeV}$
using a nonrelativistic $b$ quark and the MILC lattices described
above.\cite{Wingate:2003gm}
As with $f_B$, the perturbative matching is the dominant source of 
systematic uncertainty.  This uncertainty largely cancels in the
ratio
${f_{B_s}}/{f_{B}} = 1.20 \pm 0.03 \pm 0.01$,
where the first uncertainty is statistical error in the correlation
functions and in the chiral extrapolation, and the second uncertainty
estimates truncated terms in the usual expansions.\cite{Gray:2005ad}
Combining this ratio with the JLQCD ratio of $\mathrm{B}$ 
factors\cite{Aoki:2003xb} yields\cite{Okamoto:2005zg}
\beq
\frac{f_{B_s}}{f_{B}}\sqrt{\frac{\hat\mathrm{B}_{B_s}}{\hat\mathrm{B}_B}}
~=~ 1.210 ~{}^{+47}_{-35} \, .
\label{eq:xi}
\eeq
Ratios of the $B$ and $B_s$ decay constants and B factors have also been
determined in the static limit with 2 flavors of sea quarks
using the domain wall fermion action.\cite{Gadiyak:2005ea}

The constraint obtained by combining the improved LQCD ratio
(\ref{eq:xi}) with the new CDF measurement\cite{CDF:fpcp2006} 
(and $\lambda = 0.225(1)$\cite{Blucher:2005dc})
more than halves the width of the allowed annulus
to $\sqrt{(1-\bar\rho)^2 + \bar\eta^2}=0.93(3)$, as
shown in Fig.~\ref{fig:Vtdconstraint}b. 


The $B_s^0 - \overline{B_s^0}$ lifetime difference is also an
interesting quantity, but one for which theoretical progress
is desirable.\cite{Lenz:2004nx}  Lattice QCD can contribute
by calculating $\langle \overline{B_s} | (\bar{b} s)_{S-P} 
(\bar{b} s)_{S-P} | B_s\rangle$.  A preliminary result from
JLQCD was reported some time ago: the ratio of the correct
matrix element to the vacuum-saturation approximation was found to
be $\mathrm{B}_{S}(m_b) = 0.86(3)(7)$ in the $\overline{\rm MS}$
scheme.\cite{Yamada:2001xp}

\section{Penguins and Associated Challenges}
\label{sec:penguin}

\subsection{Rare $B$ Decays}

Lattice QCD could provide input to tests of the CKM model
using radiative decays $B\to V\gamma$.
Belle has measured branching fractions for 
$B\to (\rho/\omega) \gamma$.\cite{Abe:2005rj}
Many new flavor models predict enhanced $B\to K^*\gamma$ decays.
While LQCD input for the relevant form factor is highly 
desirable, serious technical problems must first be 
solved.\cite{Becirevic:2002zp}
At the physical point, $q^2=0$, calculations done
in the $B$ rest frame have a final state hadron with momentum
much greater than the inverse lattice spacing, ensuring lattice
artifacts dominate.  
The challenge in working with small $q^2$ with present methods,
i.e.\ in the $B$ rest frame, is that lattice artifacts dominate
when the light meson's momentum becomes comparable to the 
inverse lattice spacing.  Efforts are underway to
formulate and employ lattice NRQCD in a frame where the $B$ is moving, 
thus keeping the light meson's momentum small compared to the inverse 
lattice spacing.\cite{Hashimoto:1995in,Sloan:1997fc,Foley:2002qv,Foley:2005fx,Dougall:2005zh}  
In addition, one again has to face questions regarding extrapolating
lattice data through thresholds.

Lattice QCD can play a more immediate role in rare $B$ decays by 
calculating the form factors for $B\to K^{(*)} \ell^+\ell^-$ decays.
Branching fractions apparently agree with the Standard Model
predictions,\cite{Aubert:2003cm,Ishikawa:2003cp,Aubert:2006vb} 
but these predictions rely on
sum rule determinations for the form factors.  As the experimental
precision increases, the burden will turn to LQCD to reduce 
theoretical uncertainty.
In fact, the LQCD calculation for the most important $B\to K\ell^+\ell^-$
matrix element (within the Standard Model),
$\langle K| \bar s\gamma_\mu b |B\rangle$,
is very similar to the $B\to\pi\ell\nu$ form factor calculations discussed
above (Sec.~\ref{sec:sl}).\cite{Ali:1999mm}  
The only difference is that the mass of the strange quark 
should be held fixed as the mass of the spectator light quark is 
chirally extrapolated.  $B\to K^*\ell^+\ell^-$ would be worth
exploring as well.

\subsection{$\Delta B = 0$ Matrix Elements}

Lattice QCD can improve theoretical calculations of
the $B$ meson lifetimes through calculations of
$\langle B | {\cal O}_{\Delta B = 0} | B\rangle$,
where the ${\cal O}$ represents four 
$\Delta B = 0$ 4-quark operators.\cite{Neubert:1996we} 
Within the context of a certain flavor physics model,
e.g.\ the CKM model, the experimental $B$ meson (and $\Lambda_b$)
lifetimes can be used to test the validity of the quark-hadron
duality required to express the lifetimes in terms of those
matrix elements.
Two main challenges obstruct this calculation; these are familiar
from LQCD calculations of 
$\langle \pi| {\cal O}_{\Delta S = 1} |K \rangle$ for $K\to\pi\pi$.
First, these operators mix with lower dimensional ones, and a power-law
subtraction must be performed.  Second, it is very difficult technically
to calculate the contribution to the matrix
element when the light quark and antiquark in ${\cal O}(x)$ are 
contracted to form a propagator $G(x,x)$.
These challenges are not insurmountable and deserve to be met.
An exploratory study was reported some time ago.\cite{Becirevic:2001fy}

These calculations would have impact beyond the $B$ meson
lifetimes themselves.  Again using quark-hadron duality and
the optical theorem, LQCD calculations of
the matrix elements of two $\Delta B= 0$ 4-quark operators would be helpful
in improving the determination of $|V_{ub}|$ from 
$\overline{B}\to X_u \ell^-\bar\nu$.\cite{Voloshin:2001xi}

\section{Charmed Beauty}
\label{sec:Bc}

Using the MILC configurations discussed above, a very precise 
prediction for the $B_c$ mass was recently obtained:
$m_{B_c}^\mathrm{latt} = 6304\pm 12\,{}^{+18}_{-0}$ 
MeV.\cite{Allison:2004be}  This lattice QCD prediction has
since been confirmed by experiment, 
$m_{B_c}^\mathrm{expt} = 6287.0\pm 4.8 \pm 1.1$ MeV.\cite{Acosta:2005us}
The correlation function necessary for the calculation used
a nonrelativistic action for the $b$ quark and the Fermilab 
relativistic action for the $c$ quark.  Consequently, the agreement 
between LQCD and experiment further supports the use of both heavy 
quark formulations and improved staggered sea quarks.

\section{Conclusions}
\label{sec:concl}

The past 2 years have been especially fruitful for the interplay
between lattice QCD and heavy flavor physics, with sharp
improvements in LQCD and experimental results.  Precise measurements
of $B\to\pi\ell\nu$ decay combined with LQCD calculations of 
$B\to\pi$ form factors are making exclusive determinations 
of $|V_{ub}|$ as precise as inclusive determinations, with prospects
for further improvement.  Employing
an improved staggered quark action, LQCD calculations can 
be done with light quark masses inside the chiral regime.
The resulting prediction of $f_B$ has since been confirmed
with the first measurement of $B\to\tau\nu$ decay.
Furthermore, removing chiral extrapolation and quenching
errors from the ratio $f_{B_s}/f_B$ combines with the
first measurements of $\Delta m_s$ to drastically 
improve the precision of $|V_{td}|$.  The orthogonality of
the $|V_{td}|$ and $\sin 2\beta$ constraints on $(\bar\rho,\bar\eta)$
creates a small target for other CKM constraints to hit.  
Most of us hope the Cabibbo-Kobayashi-Maskawa trio's aim
will eventually fail, cracking the Standard Model.  If not, 
the burden is on the model builders to explain the remarkable
craftsmanship of the CKM mechanism.

\section*{Acknowledgments}

The work was supported by DOE grant number DE-FG02-00ER41132.
I am grateful to my collaborators C.\ T.\ H.\ Davies,
A.\ Gray, E.\ Gulez, G.\ P.\ Lepage, and J.\ Shigemitsu.
I thank L.\ Lellouch, C.-J.\ D.\ Lin, S.\ Sharpe 
and A.\ Soni for discussions.


\end{document}